# Healthy sweet inhibitor of Plasmodium falciparum aquaglyceroporin

**Liao Y. Chen**

Department of Physics, University of Texas at San Antonio, One UTSA Circle, San Antonio, Texas 78249 USA

Email: Liao.Chen@utsa.edu

Tel: (210)458-5457

Fax: (210)458-4919

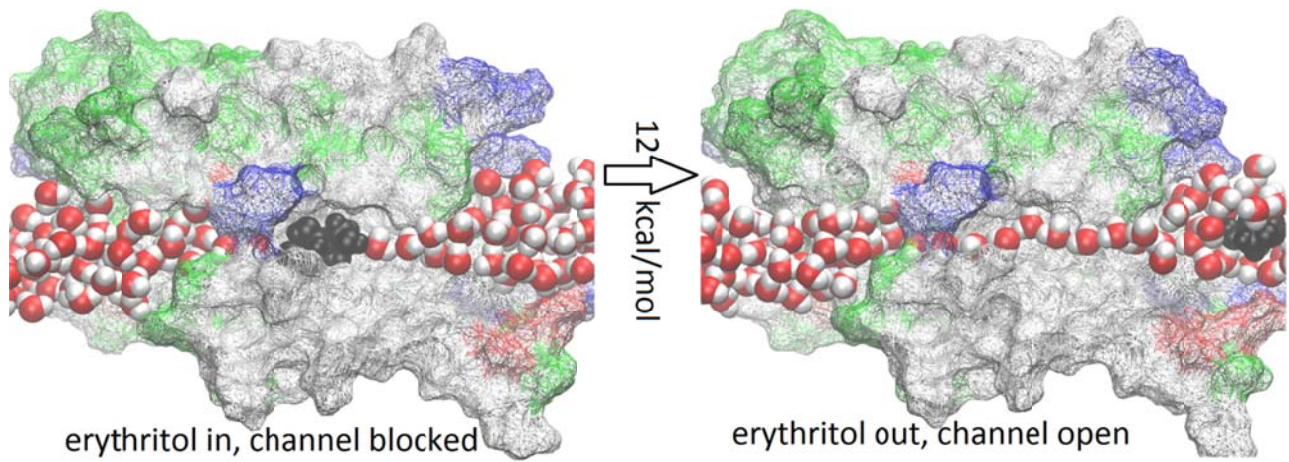

erythritol in, channel blocked — 12 kcal/mol — erythritol out, channel open

**ABSTRACT**

Plasmodium falciparum aquaglyceroporin (PfAQP) is a multifunctional channel protein in the plasma membrane of the malarial parasite that causes the most severe form of malaria infecting more than a million people a year. Finding a novel way to inhibit PfAQP, I conducted 3+ microseconds *in silico* experiments of an atomistic model of the PfAQP-membrane system and computed the chemical-potential profiles of six permeants (erythritol, water, glycerol, urea, ammonia, and ammonium) that can be efficiently transported across P. falciparum's plasma membrane through PfAQP's conducting pore. The profiles show that, with all the existent *in vitro* data being supportive, erythritol, a permeant of PfAQP itself having a deep ditch in its permeation passageway, strongly inhibits PfAQP's functions of transporting water, glycerol, urea, ammonia, and ammonium (The $IC_{50}$ is in the range of high nanomolars). This suggests the possibility that erythritol, a sweetener generally considered safe, may be the drug needed to kill the malarial parasite *in vivo* without causing serious side effects.

**INTRODUCTION**

*Plasmodium falciparum* aquaglyceroporin (PfAQP), a member of the aquaporin family[1-6], is a multifunctional channel protein on the plasma membrane of the malarial parasite that is responsible for the most severe form of malaria infecting over a million people a year. Now we have learnt from the the functional experiments that PfAQP facilitates permeation of water, glycerol, erythritol, urea, ammonia, and, possibly, ammonium across the cell membrane[7-12]. Is it possible for us to inhibit some of PfAQP's essential functions with one of its permeants? An affirmative answer to this question would open a new way of inhibiting this and other multifunctional channel proteins. So far, *in vitro* experiments supplied us with unostentatious but unambiguous evidence that glycerol inhibits water permeation through PfAQP.[7, 9, 13] The crystallization experiments aided by *in silico* simulations gave us the atomistic details of this and other aquaporin proteins, illustrating how waters and glycerols line up in a single file inside the conducting pore of this aquaglyceroporin and the Escherichia coli aquaglyceroporin GlpF[9, 14-16]. In absence of glycerol, *in vitro* experiments showed that water easily traverses the conducting pore of PfAQP[7] and *in silico* studies corroborated it with a flat landscape of its free energy[15, 17, 18]. In presence of glycerol, *in vitro* data showed reduced water permeability of PfAQP[9, 19] and *in silico* simulations produced a chemical-potential profile of glycerol having a ditch[20] in its permeation path through the protein. Glycerol, when permeating through the protein channel, would dwell inside the protein for a significant time like being in a bound state and thus occlude the conducting pore of the protein. The existence of such a ditch along glycerol's permeation path is due to the structural fitness of PfAQP hosting a glycerol near the channel center where the van der Waals (vdW) interactions are all attractive between a glycerol and the lumen residues of an aguaglyceropoin. In light of these understandings, I hypothesize that erythritol, a permeant having an even better fit than glycerol in the center of the PfAQP channel, would have a deeper ditch in its permeation path and thus would strongly inhibit PfAQP's functions of transporting water, glycerol, and other permeants. To theoretically validate this hypothesis, we need an accurate determination of the chemical-potential profile for each permeant--- the three-dimensional (3D) potential of mean force (PMF)[21-25] as a function of the permeant center-of-mass (COM) coordinates along its permeation path leading from the periplasm to the entry vestibule of PfAQP, through the channel, to the cytoplasm. These chemical-potential profiles, considered on the basis of the structure information available in the literature, can ascertain the conclusion that erythritol inhibits PfAQP in the concentration range of high nanomolars.

In this paper, I present the chemical-potential profiles for the six permeants (erythritol, water, glycerol, urea, ammonia, and ammonium) computed on the basis of 3+ µs molecular dynamics (MD) simulations, which amount to more than ten times the computational efforts invested on an aquaglyceroporin in the current literature. The chemical-potential profile of each of the permeants is computed as the 3D PMF along the most probable path from the extracellular bulk, through the PfAQP channel, to the cytoplasmic bulk. The results are shown in Fig. 1. Erythritol is found to have the strongest binding with PfAQP. It's chemical-potential curve has a very deep well inside the PfAQP channel near the Asn-Leu-Ala (NLA) and Asn-Pro-Ser (NPS) motifs that is 12 kcal/mol below its chemical potential in the bulk of periplasm/cytoplasm, leading to a dissociation constant of 786 nM. When an erythritol molecule is dissociated from and bound to the ditch inside the protein's conducting pore, PfAQP will switch between being open and closed to permeation of water or other permeants. Therefore, we can readily regulate PfAQP's functions of transporting water and other permeants by adjusting the erythritol concentration in the nM to µM range. Even though direct validation of this theory of erythritol inhibiting PfAQP needs further *in vitro* experiments measuring the erythritol-PfAQP dissociation constant and the protein's permeabilities for various erythritol concentrations, the available *in vitro* data in the current literature all support various biophysical implications of the chemical-potential profiles presented in Fig. 1.

**METHODS**

**System setup.** This study was based on the following all-atom model of PfAQP embedded in the cell membrane (illustrated in Fig. S1). The PfAQP-membrane system with glycerols was fully described in Ref. [20]. Briefly, the PfAQP tetramer, formed from the crystal structure (PDB code: 3C02), was embedded in a patch of fully hydrated palmitoyloleylphosphatidyl-ethanolamine (POPE) bilayer. The PfAQP-POPE complex is sandwiched by two layers of water, each of which is approximately 30Å in thickness. The system is neutralized and ionized with $Na^+$ and $Cl^-$ ions at a concentration of 150 mM. The entire system, consisting of 150,459 atoms, is 116Å×114 Å×107 Å in dimension when fully equilibrated. The Cartesian coordinates are chosen such that the origin is at the geometric center of the PfAQP tetramer. The xy-plane is parallel to the lipid-water interface and the z–axis is pointing from the periplasm to the cytoplasm. The PfAQP-membrane systems with ammonium, ammonia, urea, and erythritol were, respectively, built by replacing glycerols with the corresponding permeants. For the study of water permeation, the PfAQP-membrane system was obtained by deleting all the glycerols. In the case of ammonium, the system was re-neutralized with additional $Cl^-$ ions.

All the simulations of this work were performed using NAMD 2.9 [26]. The all-atom CHARMM36 parameters [27-29] were adopted for inter- and intra-molecular interactions. Water was represented explicitly with the TIP3 model. The pressure and the temperature were maintained at 1 bar and 293.15 K, respectively. The Langevin damping coefficient was chosen to be 5/ps. The periodic boundary conditions were applied to all three dimensions, and the particle mesh Ewald was used for the long-range electrostatic interactions. Covalent bonds of hydrogens were fixed to their equilibrium length. The time step of 1 fs was used for short-range interactions and 4 fs was used for long-range forces. The cut-off for long-range interactions was set to 12Å with a switching distance of 10Å. In all simulations, the alpha carbons on the trans-membrane helices of PfAQP within the range of −10Å<$z$<10Å were fixed to fully respect the crystal structure.

**Steered molecular dynamics (SMD).** SMD[30-32] runs were conducted to sample the transition paths of each of the permeating molecules (erythritol, water, glycerol, urea, ammonia, ammonium) going from the periplasm, through the conducting pore, to the cytoplasm, for computing the chemical-potential profiles. The entire permeation path leading from the periplasm (z<−25Å), through the PfAQP pore, to the cytoplasm bulk region (z>25Å), was divided into multiple sections of 1.0 Å each in width. In each section, a chosen permeant's COM was steered/pulled in the positive z-direction to sample a forward pulling path, and then pulled in the negative z-direction to sample a reverse pulling path. The pulling speed was v=2.5 Å /ns in both directions. Four forward paths and four reverse paths were sampled in every section. 4.0 ns equilibration was performed at both end points of each section so that the pulling paths were sampled between equilibrium states with the permeant's COM coordinates being fixed at the desired values. The pulling of a permeant's COM was implemented in the single-file vs non-single-file regions with one subtle but important difference:

In the single-file region inside the conducting pore ($z_1 < z < z_2$), the z-coordinate of the pulled molecule's COM was advanced with the constant velocity given above while the x- and y-degrees of freedom were left uncontrolled to obey the system's dynamics. Thus the x- and y-coordinates fluctuate around the free energy minimum $(x^*(z), y^*(z))$ on the xy-plane at a given z-coordinate. The COM of the pulled molecule approximately follows the most probable path $(x^*(z), y^*(z), z)$ as it is pulled in the z-direction. Along each forward pulling path from A to B, the work done to system was recorded as $W_{A \to Z}$ when a chosen permeant was pulled from A to Z. Along each reverse pulling path from B to A, the work done to system was recorded as $W_{B \to Z}$ when the chosen permeant was pulled from B to Z. Here Z represents a state of the system when the COM z-coordinate of the pulled molecule is z. A and B represent two end states of a given section, respectively. The 1D PMF of the chosen permeant, $G_{1D}(z)$, when its center of mass is at a given coordinate z, can be computed through the Brownian dynamics fluctuation-dissipation theorem (BD-FDT)[33] as follows:

$$G_{1D}(z) - G_{1D}(z_A) = -k_B T \ln\left( \frac{\langle \exp[-W_{A \to Z} / 2k_B T] \rangle_F}{\langle \exp[-W_{Z \to A} / 2k_B T] \rangle_R} \right). \tag{1}$$

Here $W_{Z \to A} = W_{B \to A} - W_{B \to Z}$ is the work done to the system for the part of a reverse path when the the chosen permeant was pulled from Z to A. $k_B$ is the Boltzmann constant and $T$ is the absolute temperature. $z_A$ and $z_B$ are the z-coordinates of the COM of the pulled permeant at the end states A and B of the system, respectively.

In the non-single-file regions from the channel entry vestibule to the extracellular bulk and from the channel exit to the cytoplasmic bulk, the z-coordinate was advanced in the same manner as in the single-file region but the x- and y-degrees of freedom were fixed to their constant values $(x^*(z_1), y^*(z_1))$ on the extracellular side and $(x^*(z_2), y^*(z_2))$ on the cytoplasmic side respectively. Recording the pulling work in the same manner as in the single-file region, we compute the 3D PMF of the pulled molecule, $G_{3D}(x, y, z)$, when its center of mass is at a given position (x,y,z):

$$G_{3D}(x, y, z) - G_{3D}(x_A, y_A, z_A) = -k_B T \ln \left( \frac{\langle \exp[-W_{A \to Z}/2k_B T] \rangle_F}{\langle \exp[-W_{Z \to A}/2k_B T] \rangle_R} \right). \quad (2)$$

This formula is valid for pulling along a straight line and for pulling along a fixed curve in general.

Overall, we have a continuously connected permeation path $(x^*(z_1), y^*(z_1), z)$ for $z < z_1$, $(x^*(z), y^*(z), z)$ for $z_1 < z < z_2$, and $(x^*(z_2), y^*(z_2), z)$ for $z_2 < z$ all the way from the extracellular bulk, through the conducting pore, to the cytoplasmic bulk. The chemical-potential profile along this permeation path will yield all the mechanistic essentials of permeation through PfAQP.

**3D PMF along the most probable path.** Connecting the single-file region ($z_1 < z < z_2$) to the non-single-regions ($z < z_1$ and $z > z_2$), we need to convert the 1D PMF in Eq. (1) into the 3D PMF. By definition, the 1D PMF is related to the 3D PMF as follows:[25, 34]

$$\begin{aligned} G_{1D}(z) &= -k_B T \ln \left( \int dxdy \exp[-G_{3D}(x, y, z)/k_B T]/A_{\text{ref.}} \right) \\ &= G_{3D}(x^*(z), y^*(z), z) - k_B T \ln(A(z)/A_{\text{ref.}}) \end{aligned} \quad (3)$$

where $A_{ref}$ and $A(z)$ are, respectively, the area for reference and the area occupied by the COM of a permeant on the plane of a given z-coordinate. $x^*(z)$ and $y^*(z)$ are the median coordinates of integration on the same plane. Assuming Gaussian approximation for the single file region ($z_1 < z < z_2$), the median coordinates of integration $x^*(z)$ and $y^*(z)$ are equal to the coordinates of the minimum of the 3D PMF on the xy-plane. Taking $A_{ref}$ as $A(z_1)$, we have the following approximation for the 3D PMF along the most probable path,

$$G_{3D}\left(x^*(z), y^*(z), z\right) = G_{1D}(z) + k_B T \ln\left(A(z)/A(z_1)\right) \tag{4}$$

where the area ratio can be evaluated by computing the determinant of the variance matrix of the COM x- and y-coordinates of the pulled molecule,

$$A(z)/A(z_1) = \left(\left|\begin{matrix} \langle \delta x^2 \rangle_z & \langle \delta x \delta y \rangle_z \\ \langle \delta x \delta y \rangle_z & \langle \delta y^2 \rangle_z \end{matrix}\right| \middle/ \left|\begin{matrix} \langle \delta x^2 \rangle_{z_1} & \langle \delta x \delta y \rangle_{z_1} \\ \langle \delta x \delta y \rangle_{z_1} & \langle \delta y^2 \rangle_{z_1} \end{matrix}\right|\right)^{1/2}. \tag{5}$$

Here $\delta x = x - x^*(z)$ and $\delta y = y - y^*(z)$ denote deviations from the most probable path. The brackets with subscript z mean the statistical average on the xy-plane of a given z coordinate.

**Computing the dissociation constant.** Following Ref.[24], we have the binding constant

$$\begin{aligned} c_0/k_d &= c_0 \iiint dxdydz \exp\left[-G_{3D}[x,y,z]/k_B T\right] \\ &= c_0 A(z_1) \int_{z_1}^{z_2} dz \exp\left[-G_{1D}[z]/k_B T\right]. \end{aligned} \tag{6}$$

In Eq.(6), $c_0 = 1M = 6.02 \times 10^{-4} \text{Å}^{-3}$ is the standard concentration. $k_d$ is the dissociation constant. The PMF in the bulk is chosen to be zero. For the estimations in this paper, we approximate the 3D integration of Eq. (6) as Gaussian around the minima of each permeant's chemical-potential curve so that the dissociation constant of each permeant is determined in terms of the depths of its 3D PMF ditches and the deviations of its COM coordinates:

$$k_d = c_0 \Bigg/ \sum_i c_0 e^{-\frac{G_{3D}[x_i,y_i,z_i]}{k_B T}} \begin{vmatrix} \langle \delta x^2 \rangle_i & \langle \delta x \delta y \rangle_i & \langle \delta x \delta z \rangle_i \\ \langle \delta x \delta y \rangle_i & \langle \delta y^2 \rangle_i & \langle \delta y \delta z \rangle_i \\ \langle \delta x \delta z \rangle_i & \langle \delta y \delta z \rangle_i & \langle \delta z^2 \rangle_i \end{vmatrix}^{1/2}. \tag{7}$$

In the numerator of Eq. (7), we use the standard concentration $c_0 = 1M$ while in the denominator we use its equivalent $c_0 = 6.02 \times 10^{-4} \text{Å}^{-3}$. In this way, Eq. (7) gives the dissociation constant in molars (*M*). The brackets with subscript *i* mean the mean square deviation of the permeant's COM coordinates around the *i*-th minimum of its chemical-potential curve. These deviations are shown the supplementary figures S2 to S6 for erythritol, urea, and glycerol respectively.

**RESULTS AND DISCUSSION**

Shown in Fig. 1 are chemical-potential profiles (from top panel down) for ammonium, ammonia, water, urea, glycerol, and erythritol, respectively, each of which is the 3D PMF (kcal/mol) as a function of the permeant's COM z-coordinate ($\text{Å}$). Note that, in the single file channel region ($-10\text{Å} = z_1 < z < z_2 = 10\text{Å}$), the 3D PMF curves are along the most probable path ($x^*(z), y^*(z), z$) while outside the channel in the bulk regions ($z < z_1$ and $z > z_2$), the PMF curves are along two lines ($x^*(z_1), y^*(z_1), z$) and ($x^*(z_2), y^*(z_2), z$) respectively. For all of the six permeants, the chemical potential in the extracellular bulk is, within the margin of error, equal to that in the cytoplasmic bulk, indicating validity of the computation. These six curves cannot be directly confirmed experimentally. However, their biophysical implications are in full agreement with the existent *in vitro* data available in the current literature. Their validation can be furthered with more experiments measuring the binding affinities and the permeation regulations.

The top panel of Fig. 1 shows that ammonium permeation through PfAQP has a barrier of about 14 kcal/mol. This means that PfAQP does not effectively facilitate the transport of ammonium and, speculatively, other monovalent ions. In this regard, it is in line with other members of the aquaporin family[35]. Interestingly, the 3D PMF curve for ammonia permeation is rather shallow (Fig. 1, second panel). Its deepest well is -2.5 kcal/mol and its highest barrier is only 1.5 kcal/mol, indicating that ammonia can traverse the conducting pore of PfAQP nearly as easily as through a bulk region outside the protein. Therefore, it can be concluded that PfAQP facilitates ammonia transport across the plasma membrane of P. falciparum at a very high efficiency. Inhibiting this function of PfAQP may severely harm the parasite's metabolism.

The third panel of Fig. 1 shows the 3D PMF for water permeation that has multiple shallow minima and low maxima, giving an Arrhenius activation barrier of 3 kcal/mol for water permeation. This profile indicating high water permeability is in full agreement with the *in vitro* findings [7, 9] that PfAQP is as efficient as the water specific aquaporins AQP1 and AQPZ. Inhibiting this extremely efficient function in water permeation would seriously reduce the parasite's ability of maintaining hydro-homeostasis across its plasma membrane when it is subject to severe osmotic stress during renal circulations. Inside the PfAQP channel, there are seven minima whose locations correspond well with the locations of waters found in the crystal structure of PfAQP (Fig. 2, top right, bottom left, and bottom right illustrations), which is similar to AQPZ[36]. Also, the barrier value of 3 kcal/mol resembles that of AQPZ that was experimentally measured [19].

The fourth and fifth panels of Fig. 1 show that the chemical-potential profiles of urea and glycerol both have deep ditches inside the single-file channel regions. Urea and glycerol, when permeating PfAQP, are both subject to the process of being bound inside the protein (dwelling inside the channel for significant times), occluding the channel from conducting water and ammonia. The binding affinities of urea and glycerol are similar, both in the low mM range. In terms of dissociation constant, we have $k_d^{urea} = 2.9$ mM and $k_d^{gol} = 1.8$ mM, using Eq. (7),. (Note that Ref.[20] did not estimate this dissociation constant with good enough accuracy.) This leads to the prediction that, in the mM concentration range, either urea or glycerol inhibits water/ammonia permeation through PfAQP. In presence of high concentration of glycerol/urea, PfAQP will be saturated with glycerol/urea inside its channel. Then water permeation is fully correlated to the off rate of glycerol/urea and thus will have an activation barrier (over 6 kcal/mol) much higher than the barrier (around 3 kcal/mol) in absence of glycerol/urea. The current literature does not have all the data to fully validate this theoretical prediction but the existent *in vitro* experiments do show that the presence of glycerol suppresses water conduction as discussed in details in Ref.[37].

Another similarity between glycerol and urea is that the highest barriers of their chemical-potential profiles are both in the ar/r selectivity-filter region. The Arrhenius activation barriers for their transport are, respectively, $7.7 \pm 1.5$ kcal/mol for urea and $11.5 \pm 1.5$ kcal/mol for glycerol. *In vitro* experiments are needed to measure the activation barriers and to measure the dissociation constants of urea and glycerol. Note, however, that the Arrhenius activation barrier for glycerol permeation through GlpF was measured to be 9.7 kcal/mol and the same for water permeation 7 kcal/mol[19]. These two measured values validate GlpF's chemical-potential profile[37] that resembles PfAQP's profile discussed above.

The sixth panel of Fig. 1 shows the chemical-potential profile of erythritol. The activation barrier of erythritol permeation is slightly higher than that of glycerol and thus the conclusion that PfAQP is slightly less permeable to erythritol then to glycerol. This conclusion is in agreement with the *in vitro* data of Ref. [7]. One implication of this profile is that equilibrium of erythritol concentrations between the extracellular and the cytoplasmic sides can be quickly reached when erythritol is added to the host solution. Assuming that human aquaglyceroporins are permeable to erythritol (just like PfAQP), any concentration of erythritol in the blood serum would permeate the red cell membrane and the P. falciparum membrane to reach the parasite's cytoplasm. Therefore erythritol equilibrium is expected to be quick between the blood serum and the malarial parasite's cytosol once the sweetener enters into and stays in the blood stream.

There is a very deep ditch (about 12 kcal/mol below the bulk level) in erythritol's permeation path inside the PfAQP channel. The deep ditch owes its existence to the van der Waals interactions between erythritol and the lumen residues of PfAQP that are all attractive when erythritol is near the NPS-NLA motifs. This aquaglyceroporin's structure is such that it hosts an erythritol inside with perfect fitting (Fig. 2, top left illustration). Shown in Fig. 2, bottom middle panel, is the van der Waals energy between erythritol and PfAQP as a function of the z-coordinate of the erythritol COM along a representative path of traversing the PfAQP channel. These attractive interactions are responsible for the chemical-potential profile of erythritol (Fig. 1) because the hydrogen-bonding interactions among the erythritol, the lumen residues, and the waters inside the channel do not vary drastically throughout the channel. The dissociation constant of erythritol from the PfAQP pore is estimated to be $k_d^{ery} = 786$ nM, using Eq. (7). The binding affinity of erythritol is much stronger than urea or glycerol. Therefore, the presence of glycerol or urea in the system will not inhibit the permeation of erythritol through the PfAQP pore but, conversely, the presence of erythritol with concentration ranging from high nanomolars up will inhibit the transport of glycerol, urea, water, ammonia and ammonium facilitated by PfAQP. Note that what is relevant is the erythritol concentration that is generally in equilibrium between the two sides of the P. falciparum plasma membrane, not the concentration gradient that would only show up during the initial transient after erythritol is introduced into the host solution unless the parasite actually consumes erythritol. If the parasite does metabolize erythritol, depending on the rate of consumption, a concentration gradient would exist in the cytoplasmic direction. Even in such a case, the PfAQP pore will still be occupied and thus occluded by an erythritol with a probability of nearly 100% if the extracellular concentration of erythritol is far above the $IC_{50}$ of 786 nM. Moreover, erythritol has a long biological half-life in the human blood stream. Therefore, any amount of erythritol would practically inhibit the highly efficient functions of PfAQP in facilitating transport of water, ammonia, urea, and glycerol.

**CONCLUSIONS**

The multi-functional channel protein PfAQP is capable of facilitating transport of water, ammonia, urea, glycerol, and erythritol across the P. falciparum plasma membrane. Water and ammonia can traverse the protein's conducting pore without dwelling inside the channel much longer than occupying another space of similar volume in the bulk because their chemical-potential profiles do not have deep ditches along their permeation paths through the channel. In contrast, urea, glycerol, and erythritol all have deep ditches along their permeation paths. They all dwell inside the channel with a much greater probability than occupying a similar space in the bulk regions. Therefore, the presence of any of these three solutes in the system modulates the permeation of water and ammonia. Glycerol's ditch has a depth (-6.5 kcal/mol) similar to urea's (-6.2 kcal/mol) and, therefore, the two mutually interfere with one another's transport. Finally and most importantly, the ditch along erythritol's permeation path (-12 kcal/mol) is much deeper than those of urea and glycerol. The presence of erythritol strongly regulates PfAQP's multifunctional capabilities. These conclusions can be considered partially validated because they do not contradict any but agree with all the *in vitro* data that are available in the current literature. They are still to be fully validated with more *in vitro* studies to measure transport regulations with a range of urea, glycerol, and erythritol concentrations and to measure the dissociation constants for urea, glycerol, and erythritol.

**ACKNOWLEDGEMENTS**

The author acknowledges support from the NIH (Grants #G12MD007591 and #GM084834) and the Texas Advanced Computing Center.

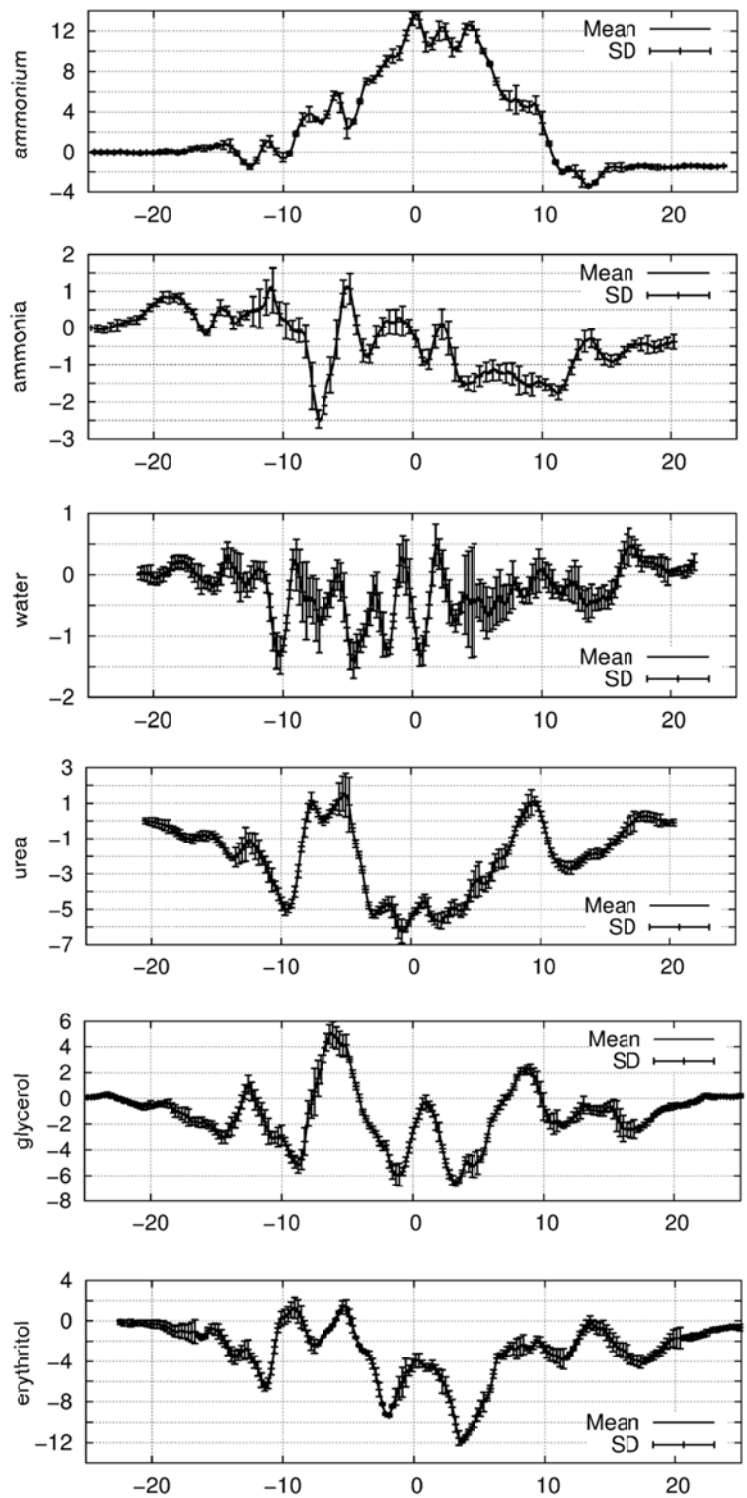

Fig.1. Chemical-potential profiles of water and solutes along the permeation passageway through the conducting pore of PfAQP: 3D PMF (kcal/mol) as a function of the z-coordinate ($\text{Å}$) of the permeating

molecule. In the bulk regions, the profiles represent the 3D PMF along two straight lines: $(x_1^*, y_1^*, z)$ leading from the extracellular bulk ($z < -10\text{Å}$) to the channel entry $(x_1^*, y_1^*, z_1 = -10\text{Å})$ and $(x_2^*, y_2^*, z)$ leading from the channel exit $(x_2^*, y_2^*, z_2 = 10\text{Å})$ to the cytoplasmic bulk ($z > 10\text{Å}$). In the single-file channel region $(z_1 < z < z_2)$, the profiles represent the 3D PMF along the most probable path $(x^*(z), y^*(z), z)$.

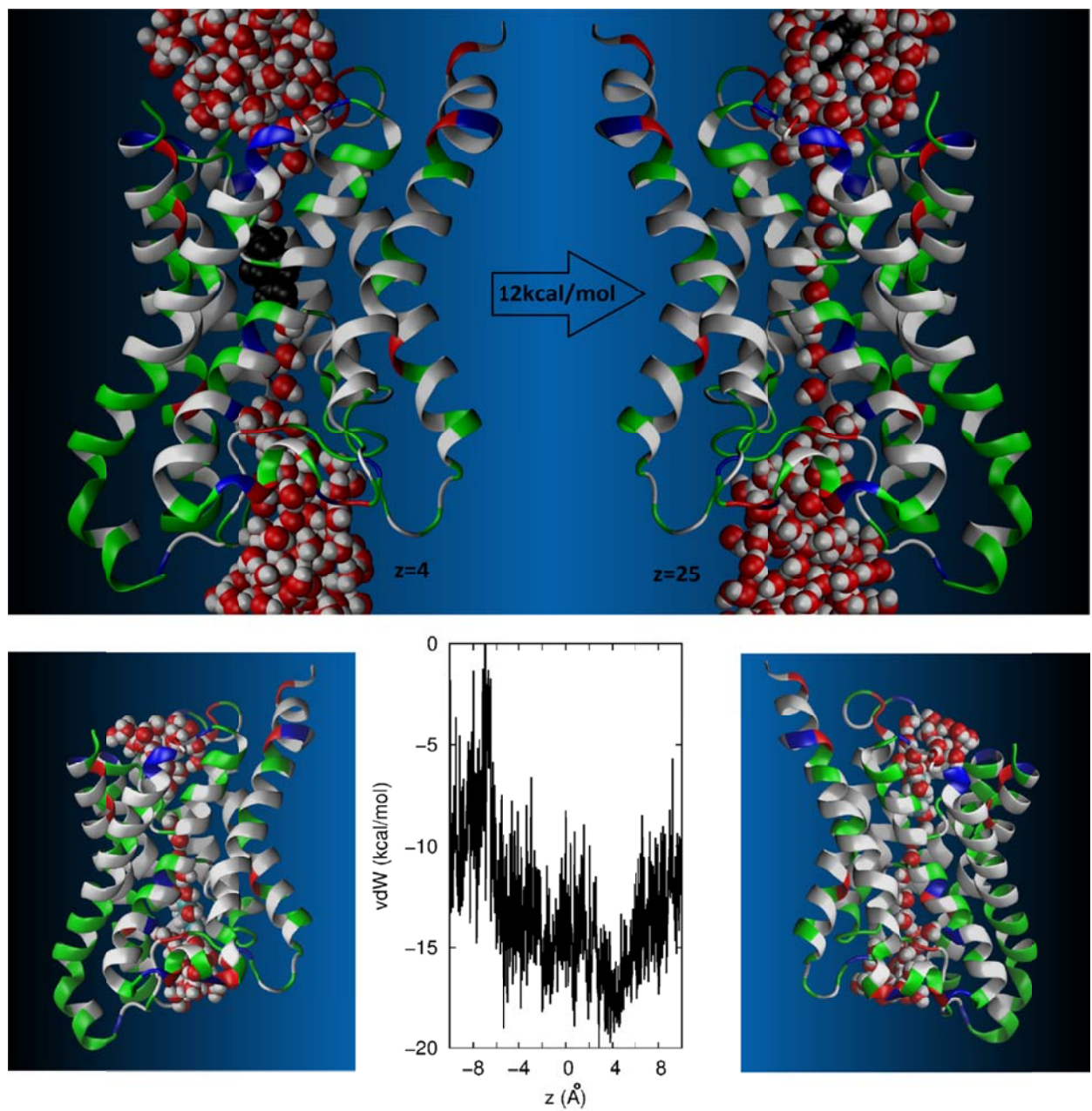

Fig. 2. Top panel, waters (in vdw, colored by element name) in and near the PfAQP conducting pore when erythritol (vdw, black) is inside the channel (left) and in the cytoplasmic bulk (right). Bottom panel, the van der Waals interaction energy between erythritol and PfAQP (central) as a function of its center-of-mass z-coordinate in the single-file region. Illustrated here are erythritol and waters (all in vdw and colored by element name) lining up inside the channel when erythritol is located at the channel entry (left) and at the channel exit (right). The protein is in newcartoon representation and colored by residue type. Graphics render with VMD[38].

## Supplementary figures to

# Healthy sweet inhibitor of Plasmodium falciparum aquaglyceroporin

Liao Y. Chen[1]

[1]Department of Physics, University of Texas at San Antonio, One UTSA Circle, San Antonio, Texas 78249 USA

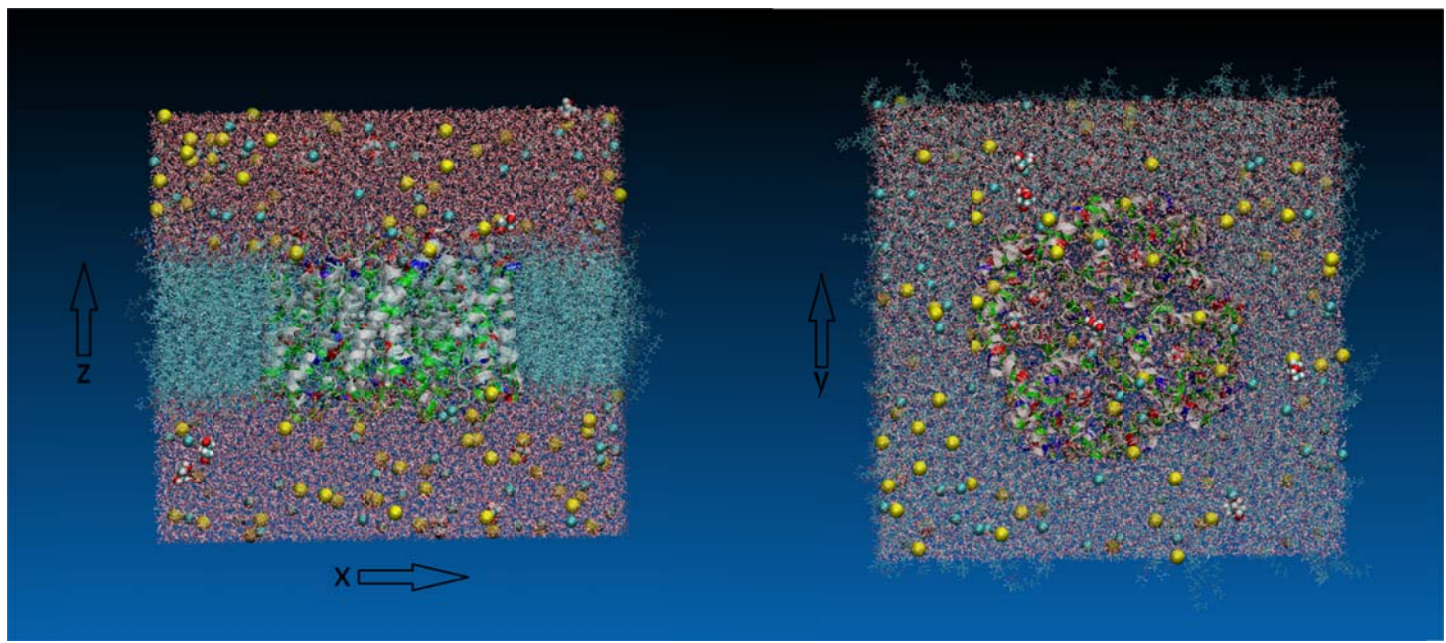

Fig. S1. All-atom model of PfAQP in the cell membrane. Shown in the left panel is the side view and in the right panel, the top view. The system is 116Å×114 Å×107 Å in dimension. Visible in the both panels are waters (in licorice representation), lipids (licorice), ions (vdw), glycerols (vdw), and the PfAQP tetramer (newcartoon). All except PfAQP are colored by element name. The PfAQP is colored by residue type. Graphics rendered with VMD (Humphrey et al., 1996).

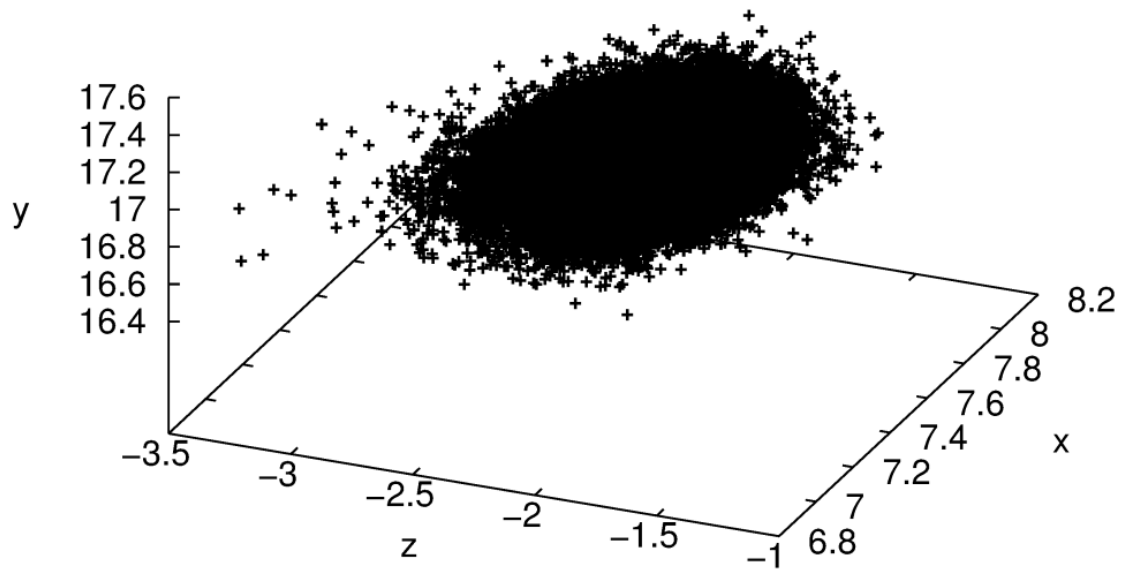

Fig. S2. The center of mass of erythritol fluctuates around the minimum at $z = -2\text{Å}$. Erythritol was free from any constraints during the equilibrium MD run of 10 ns. Its positions were sampled every 10 ps.

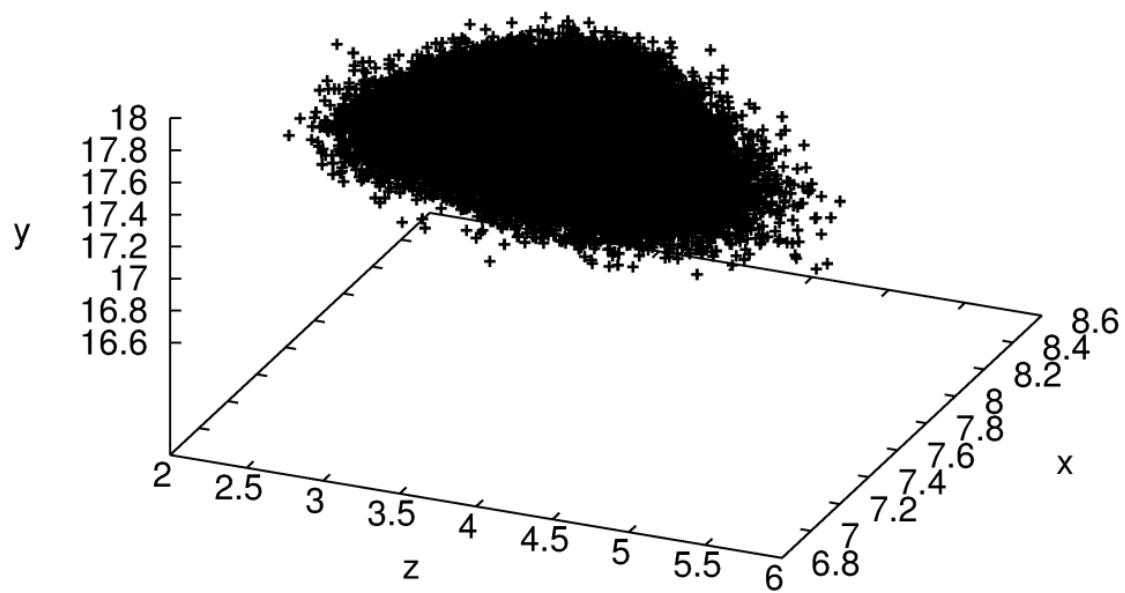

Fig. S3. The center of mass of erythritol fluctuates around the minimum at $z = 4\text{Å}$. Erythritol was free from any constraints during the equilibrium MD run of 10 ns. Its positions were sampled every 10 ps.

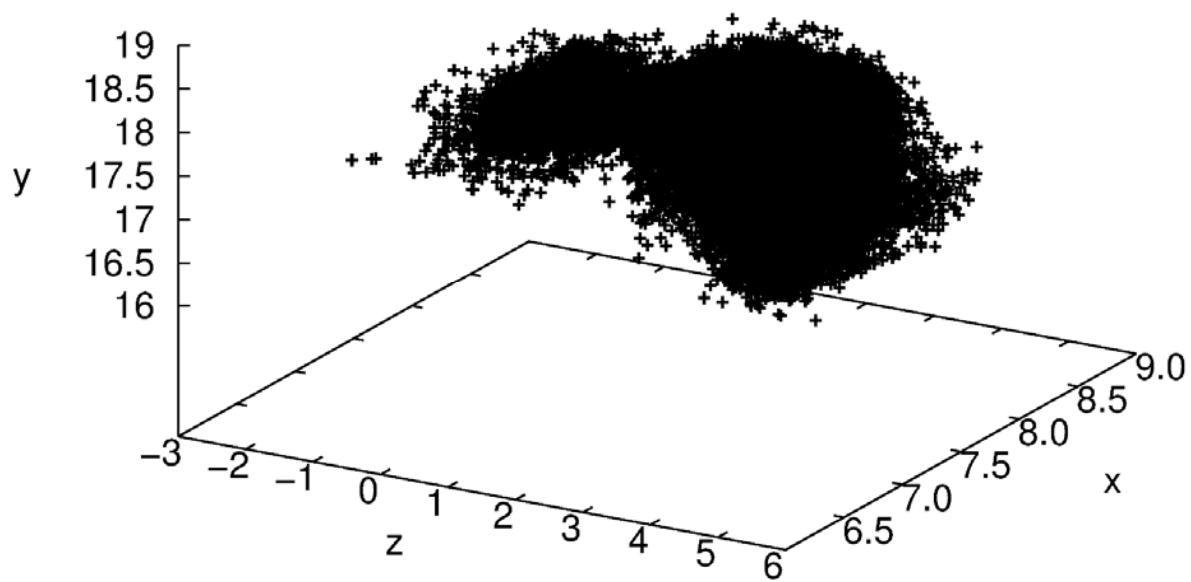

Fig. S4. The center of mass of urea fluctuates around the minima at $z = -3\text{Å}$, at $z = -1\text{Å}$, and at $z = 2\text{Å}$. Urea was free from any constraints during the equilibrium MD run of 10 ns. Its positions were sampled every 10 ps.

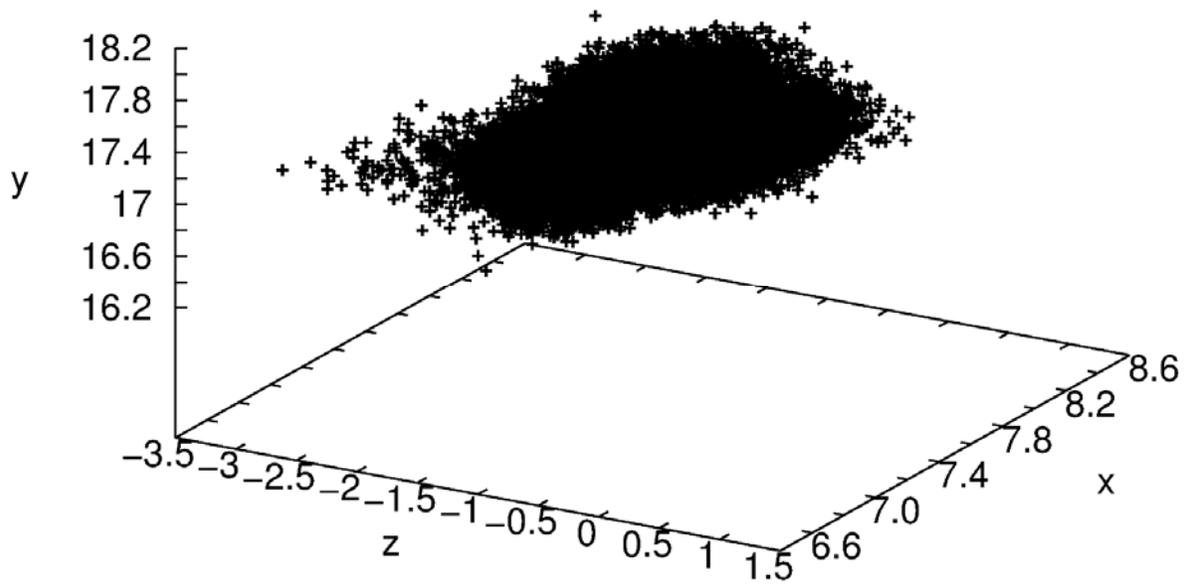

Fig. S5. The center of mass of glycerol fluctuates around the minimum at $z = -1.3 \text{Å}$. Urea was free from any constraints during the equilibrium MD run of 10 ns. Its positions were sampled every 10 ps.

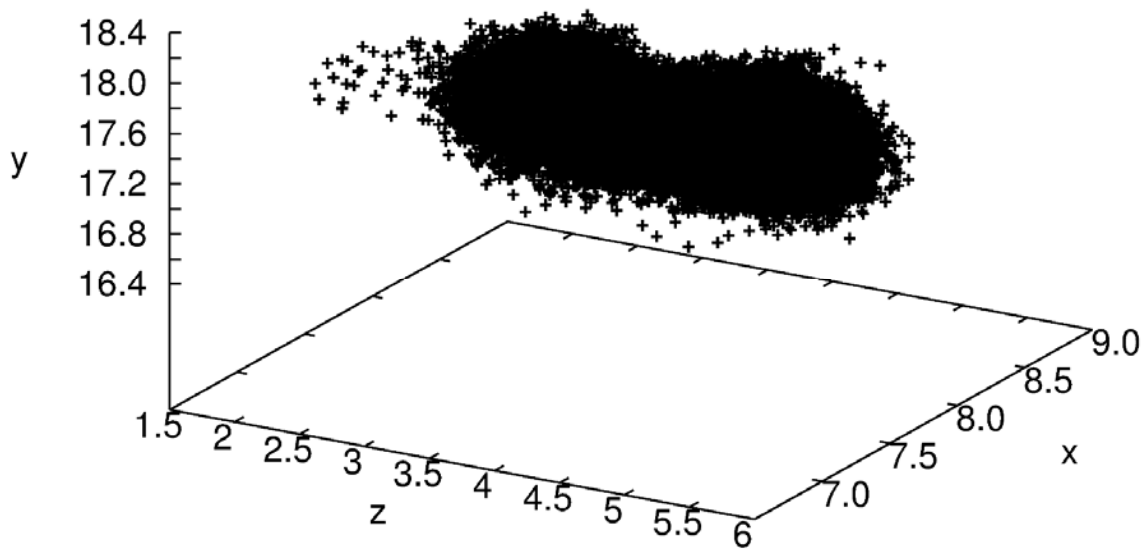

Fig. S6. The center of mass of glycerol fluctuates around the minimum at $z = 3.5\text{Å}$. Urea was free from any constraints during the equilibrium MD run of 10 ns. Its positions were sampled every 10 ps.